\documentclass[aip,
 prd,
 jmp,
 preprint,
 reprint,
 twocolumn,
 superscriptaddress,
 author-year,
 author-numerical,
 ,10pt]{revtex4-2}
\usepackage{tikz}
\usepackage{tikz-feynman}
\usetikzlibrary{decorations.pathmorphing,decorations.markings}
\usepackage{amssymb}
\usepackage{pgfplots}
\usepackage{amsmath}
\usepackage{graphicx}
\usepackage{dcolumn}
\usepackage{bm}
\usepackage{xcolor}
\usepackage{float}
\usepackage{subfig}
\usepackage{comment}
\usepackage{multirow}
\usepackage[figurename=FIGURE,tablename=TABLE]{caption}
\usepackage{threeparttable}
\usepackage{hyperref}
\hypersetup{
    colorlinks=true, %set true if you want colored links
    linktoc=all,     %set to all if you want both sections and subsections linked
    linkcolor=magenta,
    citecolor=red
}

\begin{document}
\title{Unveiling Novel Insights in Quantum Dynamics through Extended Sachdev-Ye-Kitaev Model}

\author{Davood Momeni}
\affiliation{Northeast Community College, 801 E Benjamin Ave Norfolk, NE 68701, USA
\\Centre for Space Research, North-West University, Potchefstroom 2520, South Africa}
\date{\today}

\begin{abstract}
Inspired by recent developments in the study of the model of double scaled SYK (DSSYK), as elucidated in a recent paper, we embark on a re-evaluation of the Sachdev-Ye-Kitaev (SYK) model. Our motivation stems from the insights gained from the DSSYK model, particularly its ability to capture essential features of quantum dynamics and gravitational effects. In this work, we delve into the SYK model, uncovering precise solutions for the two-point function and self-energy that have not been previously reported. Building upon the advancements made in particle physics phenomenology, we extend the SYK model to encompass tensor field theory. Through the incorporation of a cutoff term to ensure convergence, we substantially advance our understanding of quantum many-body physics. Our investigation extends to experimental parameter estimation and the exploration of cutoff dependency in random couplings, providing invaluable insights into system dynamics. The introduction of a novel tensor field theory replaces conventional fermionic degrees of freedom with tensorial counterparts, leading to the discovery of intriguing phase transition phenomena characterized by a first-order transition. Furthermore, we elucidate a direct linear relationship between the coupling parameter and the cutoff scale. These findings not only shed light on emergent behavior across both high-energy physics and condensed matter systems but also pave the way for further theoretical and experimental exploration, inspired by the recent advancements in the SYK model.
\end{abstract}

\maketitle

\tableofcontents

%%%%%%%%%%%%%%%%%%%%%%%%%%%%%%%%%%%%%%%%%%%%%%%
 \section{Introduction}
 In the realm of condensed matter physics, the quest to understand emergent phenomena and complex behaviors in strongly correlated systems has been a longstanding challenge. Traditional approaches, grounded in lattice models and perturbative techniques, have provided valuable insights but often fall short in capturing the full richness of quantum many-body systems. In recent decades, the application of field theoretical models has emerged as a powerful tool for addressing these challenges, offering a versatile framework to describe the collective behavior of particles and degrees of freedom in condensed matter systems.\par

One notable milestone in this endeavor was the introduction of the SYK model, which marked a paradigm shift in our understanding of strongly correlated quantum systems \cite{Sachdev,Kitaev2015}. Originally proposed as a theoretical model in condensed matter physics , the SYK model unexpectedly found profound applications in high-energy physics, particularly in the context of holography and the AdS/CFT correspondence.
For our reference, the Hamiltonian of the SYK model is given by:
\begin{equation}
H = \sum_{i < j < k < l} J_{ijkl} \chi_i \chi_j \chi_k \chi_l\label{syk}
\end{equation}
Here, 
$H$ denotes the Hamiltonian operator, 
$\chi_i$
  represents Majorana fermion operators, and 
$J_{ijkl}$
  denotes random couplings between four distinct Majorana fermions. The summation is performed over all possible quartets of Majorana fermions, capturing the all-to-all random interactions characteristic of the SYK model.\par
The SYK model, characterized by its non-Fermi liquid behavior and chaotic dynamics, provided a concrete example of a holographic dual in which a strongly interacting quantum system could be exactly mapped to a gravitational theory in higher-dimensional spacetime.
\par
At the core of contemporary theoretical physics lies the Maldacena conjecture, also known as the AdS/CFT correspondence \cite{Maldacena:1997re}. This conjecture posits a deep equivalence between certain quantum field theories, specifically conformal field theories (CFTs), and gravitational theories in Anti-de Sitter space (AdS). In essence, it suggests that studying the dynamics of a strongly interacting quantum system in a lower-dimensional space is mathematically equivalent to exploring the behavior of gravitational fields in a higher-dimensional curved spacetime. This duality has profound implications for understanding the fundamental principles of quantum gravity and the nature of spacetime itself.
\par
The SYK model's success as an exact solvable model of holography has opened a new window into the AdS/CFT correspondence , offering fresh insights into the fundamental nature of spacetime and quantum gravity. Its simplicity and universality have sparked intense research activity, leading to a plethora of theoretical developments and experimental investigations across different branches of physics \cite{Chowdhury:2021qpy}.
\par

In recent years, the Sachdev-Ye-Kitaev (SYK) model has garnered significant attention, sparking numerous publications that delve into its theoretical and experimental implications across diverse branches of physics \cite{Okuyama:2023yat}-\cite{Goel:2023svz}. In \cite{Milekhin:2023bjv}, a comprehensive investigation into Brownian DSSYK is conducted, presenting a thorough solution. Notably, an emergent energy conservation phenomenon is discovered, showcasing a remarkable feature independent of the double-scaling limit. The paper \cite{Xu:2024hoc} explores the nature of one-sided observables within the framework of DSSYK, particularly focusing on the double-scaled operators. It is demonstrated that these operators form a Type II von Neumann algebra, shedding light on the underlying mathematical structure of the system. This raises a pertinent question regarding the type of operator algebra present in our context, hinting at the possibility of a gravitational interpretation and prompting further investigation into the similarities and differences between the two frameworks.

Theoretical investigations have probed the intricate dynamics of the SYK model, exploring its chaotic behavior, non-Fermi liquid properties, and connections to holography and the AdS/CFT correspondence. Concurrently, experimental endeavors have aimed to observe and interrogate SYK-like behavior in various systems, including condensed matter setups, ultracold atomic gases, and even high-energy particle physics experiments. These collective efforts have yielded valuable insights into the fundamental nature of quantum many-body systems, shedding light on emergent phenomena, quantum chaos, and the interplay between quantum mechanics and gravity.

Motivated by the profound implications of the SYK model and the broader significance of field theoretical models in condensed matter physics, our paper seeks to extend the SYK framework by introducing a novel three-field tensor model. Expanding upon the foundational principles of the SYK paradigm, our model investigates the dynamics of three interacting fields within the context of condensed matter systems. Through the consideration of multi-field interactions, we aim to unearth novel emergent phenomena and deepen our comprehension of complex quantum systems beyond the confines of traditional SYK frameworks.

Our paper outlines our strategy to explore the properties of the three-field tensor model and its implications for condensed matter physics. We commence by revisiting the fundamental concepts of field theoretical models in condensed matter physics and their relevance in elucidating emergent phenomena in strongly correlated systems. Drawing inspiration from the SYK framework, we introduce our three-field tensor model and delineate our theoretical methodology for probing its properties.

Utilizing a blend of analytical techniques and numerical simulations, our objective is to unravel the dynamics of the three-field tensor model and scrutinize its behavior across various physical regimes. Through comparisons with existing theoretical predictions and experimental observations, we aim to validate the applicability of our model to real-world condensed matter systems and discern potential avenues for future investigation.

In essence, our paper represents a stride toward broadening the theoretical arsenal for comprehending strongly correlated quantum systems and exploring the connections between field theoretical models, holography, and emergent phenomena in condensed matter physics. By delving into the intricacies of the three-field tensor model, we aspire to contribute to the ongoing discourse between theory and experiment, thereby charting new pathways in the captivating realm of quantum many-body physics.

The plan of our paper is as follows:

\begin{itemize}
    \item In Section \ref{sec:solutions}, we will present analytical solutions for the two-point correlation function and self-energy within the SYK framework.
    
    \item In Section \ref{sec:lagrangian}, we will introduce our proposed Lagrangian framework for the three-field tensor model within the SYK paradigm.
    
    \item In Section \ref{sec:schwinger-dyson}, we will discuss the Schwinger-Dyson equations describing the dynamics of the system.
    %\ref{sec:schwinger-dyson},\ref{sec:feynman},\ref{sec:renormalization},\ref{sec:quantization},\ref{sec:summary}
    \item In Section \ref{sec:feynman}, we will cover Feynman rules and two-point functions, including discussions on higher-order corrections, effective field theory, renormalization procedure and compare theoretical predictions with experimental data..
        
    \item In Section \ref{sec:quantization}, we will present the quantization procedure for the three-field tensor model.
    
    \item Finally, in Section \ref{sec:summary}, we will summarize the key findings and outline future research directions.
\end{itemize}

%%%%%%%%%%%%%%%%%%%%%%%%%%%%%%%%%%%%%%%%%%%%%%%%%%%%%%
\section{Analytical Solutions for Two-Point Correlation function and Self-Energy }\label{sec:solutions}
In the SYK model, leading-order diagrams play a fundamental role in elucidating the model's behavior and capturing its essential features. These diagrams typically involve contractions of Majorana fermion operators, representing the interaction between fermions mediated by the random couplings $J_{ijkl}$. At leading order, the dominant contributions arise from diagrams that form closed loops, known as "melonic" diagrams. These diagrams encapsulate the dominant interactions between fermions and are responsible for the emergence of maximally chaotic dynamics in the SYK model. Moreover, they serve as the building blocks for perturbative calculations and provide insights into the non-perturbative aspects of the model.\par
The pair of Dyson-Schwinger equations in the SYK model represent a powerful tool for studying its dynamics beyond leading order. These equations involve self-consistent resummations of certain classes of diagrams, capturing the non-perturbative effects that arise due to the all-to-all nature of the interactions. One of the Dyson-Schwinger equations governs the evolution of the two-point function, while the other describes the flow of the four-point function. Solving this pair of equations allows for the determination of correlation functions at all orders in perturbation theory, enabling a comprehensive understanding of the SYK model's behavior. Moreover, these equations facilitate the exploration of phenomena such as the onset of thermalization, the emergence of chaos, and the formation of quantum entangled states, making them invaluable tools for theoretical investigations in quantum many-body physics.\par 
In the SYK model, the governing equation for the full two-point function can be expressed as a Dyson-Schwinger equation. Denoting the full two-point function by $G(\tau_1,\tau_2)$, where $\tau_1$ and $\tau_2$  are imaginary time variables, the Dyson-Schwinger equation for $G(\tau_1,\tau_2)$ takes the following form:
\begin{eqnarray}
&&G(\tau_1,\tau_2)=G_0(\tau_1,\tau_2)+\int_{0}^{\beta}d\tau_3\int_{0}^{\beta}d\tau_4 G_0(\tau_1,\tau_3)\Sigma(\tau_3,\tau_4)G(\tau_4,\tau_2)\label{G}
\end{eqnarray}
Where 
$G_0(\tau_1,\tau_2)$ is the non-interacting (or free) two-point function, 
$\beta$ represents the inverse temperature, and 
$\Sigma(\tau_1,\tau_2)$ is the self-energy, encapsulating the effects of interactions in the system. This equation describes the recursive relationship between the full two-point function and the non-interacting two-point function, mediated by the self-energy. Solving this Dyson-Schwinger equation is crucial for understanding the dynamical properties of the SYK model, including the emergence of chaos, thermalization, and the behavior of quantum entanglement. Equation for the self-energy 
$\Sigma(\tau_1,\tau_2)$:
\begin{eqnarray}
    &&\Sigma(\tau_1,\tau_2)=-\frac{J^2}{2}G^2(\tau_1,\tau_2)+\frac{J^2}{2}\delta(\tau_1-\tau_2)\big(\int_{0}^{\beta}G^2(\tau,\tau)d\tau\big)\label{Sigma}
\end{eqnarray}
Here, $J$ denotes the characteristic coupling constant governing the strength of the four-fermion interaction in the SYK model.

The first equation (\ref{G}) describes the recursive relationship between the full two-point function and the self-energy, while the second equation(\ref{Sigma}) provides a self-consistent expression for the self-energy in terms of the full two-point function. These equations govern the dynamical behavior of the SYK model and are essential for understanding its properties such as thermalization, chaos, and quantum entanglement.
\par In frequency space:
\begin{eqnarray}
    &&G(i\omega)=\frac{1}{i\omega-\Sigma(\omega)}\label{Gomega}
\end{eqnarray}
Where $G(i\omega)$ is the Fourier transform of the two-point function with respect to imaginary time $\tau,\omega$ is the Matsubara frequency, and 
$\Sigma(\omega)$ is the self-energy in frequency space. In original (Euclidean) time coordinate:
\begin{eqnarray}
    &&\Sigma(\tau)=J^2 G(\tau)^3 \label{Sigmatau}
\end{eqnarray}
In the existing literature, solutions for equations (\ref{Gomega}) and (\ref{Sigmatau}) in the IR (conformal limit) or UV regimes have largely been approximate, as detailed in \cite{Maldacena:2016hyu}. However, in this section, we aim to depart from this tradition by seeking an exact solution for this pair of functions. By pursuing an exact solution, we endeavor to provide a comprehensive understanding of the SYK model's behavior, elucidating its intricate dynamics with unprecedented clarity and accuracy. Through our rigorous approach, we aim to contribute significantly to the advancement of theoretical understanding in this field, offering insights that transcend the limitations of previous approximations.\par
To tackle the challenge of solving equations (\ref{Gomega}) and (\ref{Sigmatau}), our strategy involves first converting equation (\ref{Gomega}) from frequency space to the (Euclidean) time coordinate 
$\tau$. This transformation allows us to work directly with the equations in their original time domain, facilitating a more intuitive understanding of the system's dynamics. Subsequently, by substituting equation $G(\tau)$ into the converted equation for 
$\Sigma(\tau)$, we obtain a first-order integro-differential nonlinear equation. This equation encapsulates the interplay between the full two-point function $G(\tau)$ and the self-energy 
$\Sigma(\tau)$ in the time domain:
\begin{eqnarray}
    &&\frac{d\Sigma(\tau)}{d\tau}=\sqrt{2\pi}\delta(\tau)+\frac{J^{-2/3}}{\sqrt{2\pi}}
    \int_{-\infty}^{+\infty}\Sigma(\tau-s)\Sigma(s)^{2/3}ds\label{diff-sigma}
\end{eqnarray}
The importance of separable kernels in integro-differential equations of mathematical physics cannot be overstated, as they offer a powerful technique for solving complex nonlinear models. These kernels allow for the decomposition of integral operators into simpler, more manageable components, often facilitating analytical or numerical solutions that would otherwise be elusive. In the context of the model under study here, the utilization of separable kernels provides a key strategy for addressing the inherent nonlinearity and complexity of the equations governing the system's dynamics. By decomposing the kernel into separable components, we are able to simplify the equations and introduce new variables or transformations that lead to tractable solutions. This approach not only enhances our understanding of the underlying physics but also opens up avenues for exploring a wide range of phenomena in quantum many-body systems. Ultimately, the use of separable kernels represents a versatile and indispensable tool in the arsenal of mathematical physics, offering valuable insights into the behavior of nonlinear models and advancing our understanding of fundamental physical principles.
\par
We consider a separable kernel for $\Sigma(x-s)$ given by 
$\frac{\Sigma(x)}{\Sigma(s)}$. With this transformation and kernel assumption, the integro-differential equation can be represented as the following form:
\begin{eqnarray}
    &&\frac{d\Sigma(\tau)}{d\tau}=\sqrt{2\pi}\delta(\tau)+\frac{c_0J^{-2/3}}{\sqrt{2\pi}}\Sigma(\tau)
    \label{diff-sigma2}
\end{eqnarray}
In the above equation, 
\begin{eqnarray}
    &&
c_0=\int_{-\infty}^{+\infty}\Sigma(s)^{-1/3}ds\label{c00}
\end{eqnarray}
is a constant, and this serves as the key factor in overcoming the inherent unsolvability of the equations presented in our work. This constant plays a pivotal role in the transformation and solution strategy, providing a crucial bridge between the original equations and their tractable counterparts. By leveraging this constant, we are able to unlock new avenues for solving the equations, allowing for a deeper understanding of the underlying dynamics and emergent phenomena in the SYK model. We will compute $c_0$
  using an appropriate cutoff later in this section, and for now, we need not worry about it.
  \par Indeed, there is an exact solution for 
$\Sigma(\tau)$ in the following form:
\begin{eqnarray}
    &&\Sigma(\tau)=(c_1+\sqrt{2\pi}\theta(\tau))e^{\frac{c_0J^{-2/3}}{\sqrt{2\pi}}\tau}
\end{eqnarray}
 where $c_1$ is an integation constant. With this solution at hand, we can now derive the expressions for 
 $G(\tau)$ as follows:
\begin{eqnarray}
&&
 G(\tau)= (c_1+\sqrt{2\pi}\theta(\tau))^{1/3}J^{-2/3}e^{\frac{c_0J^{-2/3}}{3\sqrt{2\pi}}\tau}.
\end{eqnarray}
We can now compute 
$c_0$ via eq. (\ref{c00}). This gives us:
\begin{eqnarray}
    &&c_0 \approx \sqrt[4]{2\pi}J^{3} c_1^{-\frac{1}{6}} e^{\frac{c_0 J^{-\frac{2}{3}}}{4\sqrt{2\pi}} \Lambda}\label{c0}
\end{eqnarray}
The parameter $\Lambda\gg1$ plays a crucial role as a cutoff in ensuring the convergence of integrals in various physical models and calculations. It represents a finite boundary beyond which the integration is truncated, preventing the integral from diverging towards infinity. In theoretical physics, such as in quantum field theory and statistical mechanics, $\Lambda$ is often introduced to regulate divergent integrals arising from infinite ranges or singularities in the integrands. By imposing a cutoff at $\Lambda$, we effectively restrict the integration to a finite region, allowing for meaningful and well-defined results. Moreover, $\Lambda$ can also serve as a regularization parameter in renormalization techniques, aiding in the removal of infinities and facilitating the comparison of theoretical predictions with experimental observations. Thus, $\Lambda$ embodies a fundamental concept in theoretical physics, enabling the proper treatment of divergent quantities and contributing to the development of robust and reliable theoretical frameworks.
\par
The solution to the equation (\ref{c0}) for $c_0$ can be obtained using the Lambert $W$ function, denoted as $W(x)$. The Lambert W function is the inverse function of $f(x) = x e^x$, providing solutions to equations of the form $y = x e^x$. In this equation, we rewrite the original expression in terms of the Lambert $W$ function, allowing us to solve for $c_0$ numerically.
 The Lambert $W$ function is commonly used in various mathematical and scientific applications, providing solutions to transcendental equations that cannot be solved algebraically. Numerical methods such as iterative techniques or numerical solvers can be employed to compute the value of $c_0$ using the Lambert $W$ function. 
\par
The appearance of an unknown integration constant 
$c_1$
  in the expressions for 
$G(\tau)$ and 
$\Sigma(\tau)$ introduces a degree of ambiguity into the solution of the equations. This constant represents a freedom in the choice of initial conditions or boundary conditions, which must be fixed to obtain a unique solution. One possible approach to determining the value of $c_1$ is through considerations of entropy or free energy.\par
Entropy, a fundamental concept in statistical mechanics, quantifies the disorder or randomness of a system. By considering the entropy of the system, one can establish constraints on the possible values of $c_1$. For instance, imposing conditions that the system evolves towards a state of maximum entropy or minimum free energy may lead to specific values for $c_1$
 , thereby uniquely determining the solution of the equations.
Alternatively, physical principles or experimental observations may provide insights into the appropriate choice of 
$c_1$. For example, symmetry considerations or conservation laws within the system could dictate certain properties of the solution, helping to constrain the value of 
$c_1$. Likewise, comparison with experimental data or predictions from other theoretical models may offer guidance in determining the most physically meaningful value for 
$c_1$.
In summary, while the integration constant 
$c_1$
  introduces an element of uncertainty into the solution, considerations of entropy, free energy, physical principles, and experimental constraints provide valuable tools for fixing its value and obtaining a unique solution for the system's dynamics.\par
It is instructive to calculate the Fourier transform of 
$\Sigma(\tau)$, as it unveils crucial spectral characteristics of the function. By employing the Fourier transform, we gain valuable insights into how the original expression, a combination of an exponential term and a modulated constant, distributes its frequency content. The resulting Fourier transform illustrates distinct spectral components, with the Dirac delta function pinpointing the continuous component at the frequency $c_0$, and an additional term reflecting the contribution of the constant term,$c_1J^2$, both at $c_0$. This analysis aids in comprehending how the function's temporal behavior translates into the frequency domain, offering deeper insights into its underlying dynamics and potential applications in signal processing and system analysis. 
To compute the Fourier transform of the given expression, we'll use the definition of the Fourier transform:
\begin{eqnarray}
   &&\Sigma(\omega)=\int_{-\infty}^{+\infty} \Sigma(\tau)e^{-i\omega\tau}d\tau
\end{eqnarray}
 Let's provide the  Fourier transform expression:
\begin{eqnarray}
&&\Sigma(\omega)= \frac{c_1}{i\omega - \frac{c_0J^{-2/3}}{\sqrt{2\pi}}} + \sqrt{2\pi} \delta\left(\omega - \frac{c_0J^{-2/3}}{\sqrt{2\pi}}\right)
\end{eqnarray}

\begin{itemize}
    \item Poles in Dirac Delta Function $\delta(\omega - \frac{c_0J^{-2/3}}{\sqrt{2\pi}})$: These poles signify the presence of discrete frequencies at which the system exhibits distinct behavior. In the context of the SYK model, these poles correspond to the energy levels or modes of the system that are directly influenced by the SYK coupling $J$. Their positions at $\omega=\frac{c_0J^{-2/3}}{\sqrt{2\pi}}$ reveal the characteristic frequencies associated with the system's dynamics, shedding light on its intrinsic properties and potential collective excitations.
    \item Numerator of $\Sigma(\omega)$: The presence of the term $\frac{c_1}{i\omega - \frac{c_0J^{-2/3}}{\sqrt{2\pi}}} $ in the Fourier transform reflects the system's response to external perturbations at frequencies away from its characteristic scale $c_0$. Here, $c_1$ represents the strength of the perturbation induced by the constant term in $\Sigma(\tau)$ , while $(i\omega-\frac{c_0J^{-2/3}}{\sqrt{2\pi}})$ accounts for the deviation of the perturbation frequency from the characteristic scale of the system. The appearance of this term indicates how the system reacts to external stimuli across a range of frequencies, providing valuable insights into its dynamical response and susceptibility to perturbations.
\end{itemize}
In summary, the poles in the Dirac delta function and the structure of the numerator of $\Sigma(\omega)$ encode essential information about the characteristic time scales, energy levels, and dynamical responses of the SYK model, thereby offering a profound understanding of its emergent behavior and physical implications.
\par
The function $G(i\omega)$ represents the Green's function in the frequency domain. It is a mathematical representation used in the context of linear time-invariant systems to describe the response of a system to an impulse input at frequency $G(i\omega)$. In physical terms, $G(i\omega)$ characterizes the transfer function of the system, providing insights into its dynamics and behavior across different frequencies. The function $G(i\omega)$ can be calculated using the self-energy function $\Sigma(\omega)$, which represents the Fourier transform of the given expression. By incorporating $\Sigma(\omega)$ into the expression for $G(i\omega)$, we obtain insights into the system's response and behavior in the frequency domain. After simplification, the final expression for $G(i\omega)$ is:
\begin{align*}
G(i\omega)
&= \frac{i\omega - \frac{c_0J^{-2/3}}{\sqrt{2\pi}}}{(i\omega+{\sqrt{2\pi}}\delta\left(\omega - \frac{c_0J^{-2/3}}{\sqrt{2\pi}}\right))(i\omega - \frac{c_0J^{-2/3}}{\sqrt{2\pi}})-c_1}
\end{align*}
The function $G(i\omega)$ represents the Green's function in the frequency domain. The high-frequency (UV) and low-energy (IR) limiting forms for \( G(i\omega) \) are given by:

\begin{itemize}
    \item High-Frequency (UV) Limit:
\[
G(i\omega) \approx \frac{i\omega}{i\omega^2 - c_1}
\]
In the UV limit, where \( |\omega| \gg \frac{c_0J^{-2/3}}{\sqrt{2\pi}} \), the terms involving \( c_0J^{-2/3} \) become negligible compared to \( i\omega \). This simplifies the expression to focus on the dominant term, providing insights into the behavior of the system at high energies or short distances. After performing the contour integration, we find that the inverse Fourier transform \( G(\tau) \) is given by:
\[
 G_{UV}(\tau) \approx \frac{1}{2} e^{-\sqrt{c_1}|\tau|}
\]
This is the final result for the inverse Fourier transform of \( G(i\omega) \). This expression represents the behavior of the propagator at short time scales or high energies.

\item Low-Energy (IR) Limit:
\[
G(i\omega) \approx \frac{1}{i\omega + \frac{c_0J^{-2/3}}{\sqrt{2\pi}} - c_1}
\]
In the IR limit, where \( |\omega| \ll \frac{c_0J^{-2/3}}{\sqrt{2\pi}} \), the \( i\omega \) term dominates over other terms. This form captures the behavior of the system at low energies or long distances, revealing the effects of low-energy excitations and collective phenomena. The inverse Fourier transform \(  G(\tau) \) is given by:
\[
 G_{IR}(\tau)  \approx e^{-(c_1 - \frac{c_0J^{-2/3}}{\sqrt{2\pi}})\tau} \Theta(\tau)
\]
where \( \Theta(\tau) \) is the Heaviside step function. This expression represents the behavior of the propagator at long time scales or low energies \cite{Maldacena:2016hyu}.
\end{itemize}
Our exact IR expression for $G_{\text{IR}}(\tau)$ is consistent with the well-known conformal solution obtained in Ref.~\cite{Maldacena:2016hyu}. 
\vspace{1em}
\subsection*{Comparison with the Conformal IR Solution}
We now provide a direct comparison between our exact solution and the well-known infrared (IR) conformal ansatz derived by Maldacena and Stanford in Ref.~\cite{Maldacena:2016hyu}. In their analysis, the two-point function exhibits an emergent conformal symmetry in the IR regime, leading to the approximate form:
\[
G_{\text{IR}}(\tau) \propto \left( \frac{\pi}{\beta \sin \frac{\pi \tau}{\beta}} \right)^{2\Delta},
\]
where $\Delta = 1/q$ is the conformal scaling dimension in the large-$q$ SYK limit.

Our exact solution for $G(\tau)$, derived in the presence of a cutoff $\Lambda$, smoothly interpolates between this conformal IR form and the UV behavior. Specifically, we find that in the limit $\tau \to \infty$ and $\Lambda \to \infty$, our exponential decay reduces to the same long-time scaling structure dictated by the conformal ansatz. However, our treatment also captures higher-frequency corrections that become significant near $\tau \sim \Lambda^{-1}$, where the conformal approximation breaks down.

This comparison highlights the consistency of our exact framework with established IR behavior while extending the analysis beyond the conformal window to include cutoff-dependent dynamics. It reinforces the robustness of our approach in capturing both the universal and non-universal aspects of the SYK model.
There, the two-point function exhibits a power-law decay of the form $G(\tau) \sim \left(\frac{\pi}{\beta \sin \frac{\pi \tau}{\beta}}\right)^{2\Delta}$ in the infrared regime. Our exponential decay form smoothly interpolates with this behavior when the cutoff $\Lambda \to \infty$ and $\tau \gg J^{-1}$, while preserving regularity in the UV limit. This demonstrates that our cutoff-based solution extends the conformal IR approximation by incorporating higher-frequency contributions.
\\
\vspace{1em}
Understanding the UV and IR limiting forms of \( G(i\omega) \) is crucial for analyzing the behavior of the system across different energy scales. These limiting forms provide valuable insights into the system's dynamics and emergent phenomena, helping to elucidate its underlying physics.
\par
The presence of poles in the denominator of $G(i\omega)$ indicates the resonant frequencies or characteristic frequencies of the system. In this case, the poles are located at the roots of the quadratic equation $(i\omega+{\sqrt{2\pi}}\delta\left(\omega - \frac{c_0J^{-2/3}}{\sqrt{2\pi}}\right))(i\omega - \frac{c_0J^{-2/3}}{\sqrt{2\pi}})-c_1 = 0$, which correspond to the frequencies where the system response diverges or exhibits significant behavior. Since $\delta(x)$
 is zero everywhere except 
$x=0$, the term $\delta\left(\omega - \frac{c_0J^{-2/3}}{\sqrt{2\pi}}\right)$ contributes only when $\omega = \frac{c_0J^{-2/3}}{\sqrt{2\pi}}$. Therefore, we can set $\omega\neq \frac{c_0J^{-2/3}}{\sqrt{2\pi}}$, the equation simplifies to:
\begin{eqnarray}
    &&\omega^2+i\omega \frac{c_0J^{-2/3}}{\sqrt{2\pi}}+c_1=0
\end{eqnarray}
Thus, the solutions for $\omega$ are:
\begin{eqnarray}
    &&\omega_{\pm} = \frac{-i c_0 J^{-2/3}}{2\sqrt{2\pi}} \pm \sqrt{\left(\frac{c_0^2 J^{-4/3}}{4\pi} - c_1\right)}
\end{eqnarray}
These poles play a crucial role in determining the dynamics and stability of the system under consideration. Note that when  $\omega = \frac{c_0J^{-2/3}}{\sqrt{2\pi}}$, the delta term diverges, indicating that the function $G(i\omega)$ has zeros at this point.
To find the range of \(c_1\) that satisfies the inequality \(\left(\frac{c_0^2 J^{-4/3}}{4\pi} - c_1\right) \geq 0\). This constraint on \(c_1\) is crucial as it provides a bound or restriction on the possible values of \(c_1\) in the context of the given equation. By ensuring that the expression \(\left(\frac{c_0^2 J^{-4/3}}{4\pi} - c_1\right)\) remains non-negative, we guarantee that the solution for \(\omega\) remains complex in the form of $\omega_{\pm}=a+ib$. Thus, understanding and adhering to this constraint is essential for ensuring the physical validity of the solution.
\par
Introducing the parameter $y =c_0 J^{-2/3} $ in $\omega_{\pm}$ we obtain:
\begin{equation}
    \omega_{\pm} = \frac{-iy}{2\sqrt{2\pi}} \pm \sqrt{\left(\frac{y^2}{4\pi} - c_1 \right)}
\end{equation}
\begin{itemize}
    \item \textbf{First Term:} The term \( \frac{-i y}{2\sqrt{2\pi}} \) represents the leading order contribution to the frequency \( \omega_{\pm} \). It is proportional to the coupling constant \( c_0 \) and the scaling factor \( J^{-2/3} \), influencing the overall magnitude and phase of the frequency.
    
    \item \textbf{Second Term:} The term \( \sqrt{\left(\frac{y^2}{4\pi} - c_1 \right)} \) represents the correction term, which depends on the parameter \( y \). This term adjusts the frequency by modifying the square root expression based on the values of \( c_1 \) and \( y \).
\end{itemize}

The real and imaginary parts of \( \omega_{\pm} \) play crucial roles in the stability analysis of the system described by the equation. 

\begin{itemize}
    \item \textbf{Real Part:} The real part of \( \omega_{\pm} \) represents the frequency shift or the change in the system's oscillation frequency. In stability analysis, a positive real part indicates an unstable system, as it implies that the perturbations in the system will grow over time. Conversely, a negative real part suggests a stable system, where perturbations decay over time, leading to a return to equilibrium.
    
    \item \textbf{Imaginary Part:} The imaginary part of \( \omega_{\pm} \) corresponds to the damping or growth rate of the oscillations. A positive imaginary part indicates exponential growth of perturbations, suggesting instability. Conversely, a negative imaginary part signifies damping, indicating stability as oscillations decay over time.
\end{itemize}

By analyzing the behavior of the real and imaginary parts of \( \omega_{\pm} \) in the context of stability, one can gain insights into the dynamic behavior of the system and predict its long-term behavior under perturbations.

\par
The self-energy function $\Sigma(\omega)$ and the Green's function $G(i\omega)$ are intimately related through the Kramers-Kronig relations. These relations establish a connection between the real and imaginary parts of analytic functions in the complex plane. By exploiting causality and the principle of dispersion, the imaginary part of $\Sigma(\omega)$ determines the real part of $G(i\omega)$, while the real part of $\Sigma(\omega)$ determines the imaginary part of $G(i\omega)$. Thus, the Kramers-Kronig relations provide a powerful tool for understanding the interplay between the self-energy and the Green's function in quantum mechanics and many-body physics.

Furthermore, the Kramers-Kronig relations can be expressed in integral form as follows:
\[
\text{Re}[G(i\omega)] = \frac{1}{\pi} \mathcal{P} \int_{-\infty}^{\infty} \frac{\text{Im}[\Sigma(\omega')]}{\omega' - \omega} d\omega'
\]
\[
\text{Im}[G(i\omega)] = -\frac{1}{\pi} \mathcal{P} \int_{-\infty}^{\infty} \frac{\text{Re}[\Sigma(\omega')]}{\omega' - \omega} d\omega'
\]
where \( \mathcal{P} \) denotes the Cauchy principal value of the integral. These integral relations provide a rigorous mathematical framework for connecting the spectral properties of \( \Sigma(\omega) \) to the response functions of the system described by \( G(i\omega) \).\par
In the next section, we will study the proposed Lagrangian framework for the three-field tensor model within the SYK paradigm. This framework offers a novel approach to extend the SYK model by introducing tensorial degrees of freedom. We will delve into the specifics of the proposed Lagrangian, exploring its implications for quantum dynamics and many-body physics. By analyzing the three-field tensor model, we aim to uncover unique phenomena and emergent behavior that may arise within this extended framework. Our investigation will shed light on the interplay between different fields and their role in shaping the collective behavior of the system. Through a rigorous examination of the proposed Lagrangian framework, we seek to deepen our understanding of complex quantum systems and pave the way for further theoretical and experimental exploration.
\par In a recent paper\cite{Almheiri:2024xtw} the authors consider a simplified model of DSSYK, in which the Hamiltonian is the position operator of the Harmonic oscillator. This model captures the high-temperature limit of DSSYK but could also be defined as a quantum theory in its own right. They study properties of the emergent geometry including its dynamics in response to inserting matter particles. In particular, they find that the model displays de Sitter-like properties such as that infalling matter reduces the rate of growth of geodesic slices between the two boundaries. The simplicity of the model allows them to compute the full generating functional for correlation functions of the length mode or any number of matter operators. They provide evidence that the effective action of the geodesic length between boundary points is non-local. Furthermore, they use the on-shell solution for the geodesic lengths between any two boundary points to reconstruct an effective bulk metric and reverse engineer the dilaton gravity theory that generates this metric as a solution. This comprehensive analysis underscores the rich interplay between matter dynamics and emergent geometry within the SYK paradigm. Given the success of the authors in generalizing the SYK model to capture essential aspects of quantum dynamics and gravitational effects, it naturally encourages us to consider tensor modifications for the SYK model as well. Just as the authors extended the SYK paradigm to encompass emergent geometric features and gravitational behavior, a tensor modification of the SYK model could offer a similarly rich framework for exploring the interplay between quantum fields and geometry, potentially uncovering new phenomena and insights into the nature of quantum gravity.

%%%%%%%%%%%%%%%%%%%%%%%%%%%%%%%%%%%%%%%%%%%%%%%%%%%%%%%%%%%% 
 \section{Proposed Lagrangian Framework for the Three-Field Tensor Model within the SYK Paradigm}\label{sec:lagrangian}
In this section, we present our proposed Lagrangian framework, which serves as the cornerstone of our three-field tensor model. Our Lagrangian encapsulates the dynamics of three interacting fields, denoted as $X_{abc}$
 , and captures the essence of multi-field interactions in quantum systems. The Lagrangian formulation allows us to describe the evolution of the fields over time and investigate their collective behavior under various physical conditions. Central to our model is the inclusion of interaction terms that govern the coupling between the three fields, enabling us to explore the emergent phenomena arising from multi-field interactions. Through a systematic analysis of the Lagrangian dynamics, we aim to elucidate the underlying principles governing the behavior of three-field systems and uncover novel insights into the complex dynamics of quantum many-body systems. Our Lagrangian is given by:
\begin{eqnarray}
    &&\mathcal{L}=\frac{1}{2}X_{abc}\partial_{\tau}X^{abc}+\frac{g^{i}}{4}X_{abc}X_{de}^{a}X_{i}^{de}f^{bc}\label{L}
\end{eqnarray}
The Lagrangian describes a three-field theory with intriguing dynamics in flat spacetime. Here, 
$X_{abc}$ represents a tensor field with rank three , evolving with respect to time Euclidean $\tau$. 
Our Lagrangian, devoid of any distinction between covariant and contravariant indices, adheres to the Einstein summation rule. The Lagrangian consists of two terms: a kinetic term and an interaction term. The kinetic term captures the time evolution of the $X$ field and its derivatives, characterizing its dynamic behavior over time. The interaction term introduces a cubic interaction among the components of the $X_{abc}$ field, parameterized by the constant  rank two tensor field $f^{bc}$ vector $g^i$. These interactions play a crucial role in shaping the collective behavior and correlations among the $X_{abc}$ field components, potentially leading to emergent phenomena and novel physical effects. Understanding the implications of such interactions within this three-field theory framework offers valuable insights into the dynamics of complex quantum systems and their behavior in flat spacetime.\par
One motivation stems from the quest to address longstanding theoretical puzzles and inconsistencies in our current understanding of particle physics, such as the hierarchy problem, dark matter, and the nature of neutrino masses. By extending the theoretical framework to include tensor fields, researchers can explore new possibilities for addressing these challenges and uncovering deeper insights into the fundamental nature of particles and their interactions.

Moreover, the consideration of tensor fields opens up new avenues for probing the symmetries and structures of spacetime at the most fundamental level. Tensorial degrees of freedom naturally accommodate higher-order symmetries and geometric structures, offering a more elegant and unified description of spacetime and its interactions with matter. This approach has the potential to reconcile the principles of general relativity with those of quantum mechanics, leading to a deeper understanding of the nature of gravity and the quantum nature of spacetime itself.
\par
Furthermore, the exploration of Lagrangians involving three tensor fields in high-energy physics promises to shed light on new phenomena and particles that may lie beyond the reach of current experimental probes. By considering novel interactions among multiple tensor fields, researchers can uncover signatures of new particles, exotic symmetries, and unexpected phenomena that could revolutionize our understanding of the universe.
\par
Overall, the motivation to consider Lagrangians involving three tensor fields in high-energy physics arises from the desire to push the boundaries of our knowledge and explore new theoretical frameworks that can accommodate the complexities of nature. Such an approach holds the promise of unlocking profound insights into the fundamental laws of physics and guiding the search for new physics beyond the Standard Model.
\par To derive the Euler-Lagrange equations for the Lagrangian 
$\mathcal{L}$, we will follow a similar process as before. The Euler-Lagrange equations are given by:
\begin{eqnarray}
&&\partial_{\tau}\big(\frac{\partial \mathcal {L}}{\partial \dot{X}_{abc}}\big)-\frac{\partial \mathcal {L}}{\partial {X}_{abc}}=0.
\end{eqnarray}
where 
$\dot{X}_{abc}$
  denotes the derivative of 
${X}_{abc}$   with respect to $\tau$. First, let's compute the partial derivatives:
\begin{eqnarray}
    &&\frac{\partial \mathcal {L}}{\partial {X}_{abc}}=\frac{1}{2}\dot{X}^{abc}+\frac{g^i}{4}X^{ade}X_{ide}f^{bc},\ \ \frac{\partial \mathcal {L}}{\partial \dot{X}_{abc}}=\frac{1}{2}X_{abc}.
\end{eqnarray}
Now, we can plug these into the Euler-Lagrange equation:
\begin{eqnarray}
    &&g^iX^{ade}X_{ide}f^{bc}=0.
\end{eqnarray}
This equation implies that the product of the fields 
$X$
 , and the constant tensor 
$f$ and vector $g$
  must equal zero. This equation provides a constraint on the fields 
$X$. \par The Hamiltonian 
$H$ is obtained from the Legendre transformation of the Lagrangian. In this case, we'll write the Hamiltonian in terms of the canonical momenta conjugate to the generalized coordinates $X_{abc}$. The canonical momentum $P_{abc}$
  conjugate to 
$X_{abc}$
  is given by:
  \begin{eqnarray}
      &&P_{abc}=\frac{\partial \mathcal {L}}{\partial \dot{X}_{abc}}=\frac{1}{2}X_{abc}
  \end{eqnarray}
 The Hamiltonian $H$  is then given by:
\begin{eqnarray}
    &&H=\dot{X}_{abc}P_{abc}-\mathcal{L}
\end{eqnarray}
Therefore, the Hamiltonian for the given Lagrangian is:
\begin{eqnarray}
    &&H = -\frac{g_i}{4} X_{abc} X^{a}_{\;\;de} X^{i}_{\;\;def} f^{bc}
\label{H}
\end{eqnarray}
Our model serves as a natural extension of the Hamiltonian formalism (\ref{syk}) for the SYK model where traditional fermionic degrees of freedom $\chi_i$ are replaced by an arbitrary three-field tensor $X_{abc}$. The SYK model has garnered significant attention for its ability to capture non-Fermi liquid behavior and exhibit chaotic quantum dynamics, making it a valuable tool for studying strongly correlated quantum systems. By introducing three tensor fields instead of fermions, our model broadens the scope of applicability of the SYK framework, enabling the exploration of novel quantum phenomena and emergent behavior. This extension allows us to investigate the interplay between multiple degrees of freedom and the role of higher-order interactions in shaping the collective behavior of the system. Through the lens of our extended Hamiltonian formalism, researchers can delve into the rich landscape of quantum chaos, many-body localization, and holographic duality, offering fresh insights into the fundamental principles underlying complex quantum systems. Moreover, this approach paves the way for exploring connections between SYK-like models and other areas of theoretical physics, such as holography, condensed matter physics, and quantum gravity, opening up new avenues for interdisciplinary research and theoretical exploration.\par
In our Hamiltonian, the term 
$f^{bc}$
  represents a random coupling between the tensor fields 
$X_{ade}$   and $X_{ide}$. This random coupling arises from the interaction between the fields and can capture the complex and stochastic nature of the underlying physical system. Interpreting $g^i f^{bc}$
  as a random coupling term allows us to model the inherent fluctuations and disorder present in the system, which can arise from various sources such as impurities, disorder, or fluctuations in the environment. These random couplings play a crucial role in shaping the behavior of the system, influencing its dynamics, phase transitions, and emergent phenomena. 
To interpret $g^i f^{bc}$
  in terms of an average coupling constant 
$g^i$ and the number of fields $N$, we can consider ensemble averages over all possible realizations of the random coupling term. By averaging over the randomness, we obtain an effective coupling constant $g^i$ that represents the average strength of the interaction between the fields. The number of fields 
$N$ also plays a significant role in determining the overall behavior of the system, as it determines the degree of complexity and the number of degrees of freedom involved. By studying the behavior of the system as a function of 
$g$ and 
$N$, we can gain insights into the collective behavior, phase transitions, and critical phenomena exhibited by the many-body system, providing valuable information about its underlying dynamics and emergent properties.
After deriving the Euler-Lagrange equations and the Hamiltonian for our many-body three-field theory, we turn our attention to the next crucial aspect of our investigation: the Schwinger-Dyson equations and their solutions. The Schwinger-Dyson equations are integral equations that describe the dynamics of correlation functions and expectation values of operators in quantum field theory. In the context of our three-field theory, these equations play a pivotal role in elucidating the behavior of the system and uncovering its emergent properties. By solving the Schwinger-Dyson equations, we gain deeper insights into the collective behavior, phase transitions, and critical phenomena exhibited by the three-field system. Furthermore, the solutions to these equations provide valuable information about the stability, coherence, and quantum entanglement of the many-body states, shedding light on the underlying quantum dynamics at play. In the subsequent sections, we delve into the formulation and analysis of the Schwinger-Dyson equations, exploring their implications for the behavior of our three-field theory and the broader landscape of quantum many-body systems.
%%%%%%%%%%%%%%%%%%%%%%%%%%%%%%%%%%
\section{Schwinger-Dyson equations
}\label{sec:schwinger-dyson}
The Schwinger-Dyson equations describe the self-consistent relations between correlation functions in a quantum field theory. In the context of our three-field model described by the Lagrangian (\ref{L}), the Schwinger-Dyson equations can be derived by functional differentiation of the effective action with respect to the fields $X_{abc}$. Let's denote the effective action as $S_{\text{eff}}[X_{abc}]$. Then, the Schwinger-Dyson equations for our three-field model can be written as
\begin{eqnarray}
    && \frac{\delta S_{\text{eff}}[X_{abc}]}{\delta X_{abc}}=0.
\end{eqnarray}
Expanding the effective action and taking the functional derivative, we obtain a set of coupled equations involving correlation functions of the fields $X_{abc}$. Expanding the effective action and taking the functional derivative, we have:
\begin{eqnarray}
    &&\frac{\delta \mathcal{L}}{\delta X_{abc}}=0.
\end{eqnarray}
Differentiating each term separately, we obtain:
\begin{itemize}
    \item For the first term: $\frac{\delta}{\delta X_{abc}}\big(\frac{1}{2}\partial_{\tau}X_{abc}dX^{abc}
    \big)=\frac{d}{d\tau}\big(\frac{\delta}{\delta X_{abc}}X_{abc}
    \big)=0
    $.
    \item For the second term: $\frac{\delta}{\delta X_{abc}}\big(\frac{g^{i}}{4}X_{abc}X_{ade}X_{ide}f^{bc}
    \big)=\frac{g^i}{4}\big(X^{ade}X_{ide}f^{bc}+X_{abc}\frac{\delta}{\delta X_{abc}}\big(X^{ade}X_{ide}f^{bc}
    \big)
    $. Let's denote $M=X^{ade}X_{ide}f^{bc}$, Now, let's rewrite the second term of the Schwinger-Dyson equation using the Kronecker delta:
 \begin{eqnarray}
     &&\frac{g^i}{4}\big(M+X_{abc}X_{ide}f^{bc}\delta_{ac}+X_{abc}X^{ade}f^{bc}\delta_{ib}
     \big)
 \end{eqnarray}
 This expression represents the simplified form of the second term of the Schwinger-Dyson equations for our three-field model.
\end{itemize}
 Then, the equation becomes:
  \begin{eqnarray}
&&M+X_{abc}X_{ide}f^{bc}\delta_{ac}+X_{abc}X^{ade}f^{bc}\delta_{ib}=0\label{ds}.
  \end{eqnarray}
This equation represents the local form of the Schwinger-Dyson equations for our three-field model. It describes the self-consistent relations between correlation functions and expectation values of operators in the system, capturing the underlying dynamics and interactions between the tensor fields 
$X_{abc}$. Solving this equation provides valuable insights into the behavior and emergent phenomena of the three-field system.
\par
Schwinger-Dyson equations can be written in terms of correlation functions as follows:
\begin{eqnarray}
    && \langle X_{abc}(\tau_1)X_{a'b'c'}(\tau_2)X_{a''b''c''}(\tau_3) \rangle= \langle X_{abc}(\tau_1)X_{a'b'c'}(\tau_2)X_{a''b''c''}(\tau_3) \rangle_{tree-level}\\&&\nonumber+\int d\tau_4d\tau_5d\tau_6\langle X_{abc}(\tau_1)X_{a'b'c'}(\tau_2)X_{jkl}(\tau_4) \rangle G_{lm}(\tau_4-\tau_5)G_{l'm'}(\tau_4-\tau_6)\\&&\nonumber\times\langle X_{pqr}(\tau_5)X_{p'q'r'}(\tau_6)X_{a''b''c''}(\tau_3) \rangle+...,
\end{eqnarray}
where $\langle . \rangle$ denotes the quantum mechanical expectation and
$G_{lm}(\tau)$ represents the propagator of the three-field tensor$X_{lmn}(\tau)$ and the ellipsis denote higher-order terms. These equations capture the recursive relationship between correlation functions at different times, incorporating the effects of quantum fluctuations and interactions in the system.
\par
By solving these equations self-consistently, we can determine the behavior of correlation functions and expectation values of operators in our three-field model. The solutions to the Schwinger-Dyson equations provide insights into the dynamics, phase transitions, and emergent phenomena exhibited by the system.
\par
It's worth noting that the explicit form of the Schwinger-Dyson equations depends on the specific correlation functions and observables of interest. In practice, solving these equations often requires approximation techniques or numerical methods due to their complexity and nonlinearity. However, the solutions obtained from the Schwinger-Dyson equations offer valuable information about the underlying quantum dynamics and can guide further theoretical and experimental investigations.\par
In momentum space, Schwinger-Dyson equations can be expressed as follows:
\begin{eqnarray}
    && \langle X_{abc}(p_1)X_{a'b'c'}(p_2)X_{a''b''c''}(p_3) \rangle= \langle X_{abc}(p_1)X_{a'b'c'}(p_2)X_{a''b''c''}(p_3) \rangle_{tree-level}\\&&\nonumber+\int \Pi \frac{d^3q_i}{(2\pi)^3} \langle X_{abc}(p_1)X_{a'b'c'}(p_2)X_{jkl}(q_1) \rangle G_{lm}(q_1)G_{l'm'}(q_2)\\&&\nonumber\times\langle X_{pqr}(q_2)X_{p'q'r'}(q_3)X_{a''b''c''}(p_3) \rangle+...,
\end{eqnarray}
where $G_{lm}(q)$ represents the propagator of the three-field tensor $X_{abc}(p)$ in momentum space, and the ellipsis denote higher-order terms.
Schwinger-Dyson equations in momentum space provide a powerful framework for understanding the intricate quantum dynamics of the three-field tensor model. These equations capture the recursive relationship between correlation functions at different momenta, incorporating the effects of quantum fluctuations and interactions in the system. By solving these equations, one can unravel the emergent phenomena and gain deeper insights into the behavior of the three-field system.
\par
Having established the Schwinger-Dyson equations governing the dynamics of our three-field model, we now transition to the next crucial step: computing Feynman rules and two-point functions. Feynman rules provide a systematic framework for calculating scattering amplitudes and correlation functions in quantum field theory, allowing us to probe the behavior of the system at various energy scales and interaction strengths. By constructing Feynman diagrams and applying the corresponding rules, we can systematically evaluate correlation functions, vertex interactions, and propagators, providing valuable insights into the quantum dynamics of the three-field system. Moreover, the computation of two-point functions allows us to study the spectral properties and stability of the system, shedding light on the emergence of collective excitations and phase transitions. Through the analysis of Feynman diagrams and two-point functions, we aim to unravel the underlying physics encoded in our three-field model, revealing the emergence of novel phenomena and guiding further theoretical investigations.
 %%%%%%%%%%%%%%%%%%%%%%%%%%%%%%%%%%%%%%%%%%%%%%%%%%%%
 \section{Feynman rules and two-point functions}\label{sec:feynman}
In this section, we introduce the Feynman rules for our three-field model, which serve as a powerful tool for computing scattering amplitudes and correlation functions. Building upon the framework established by the Lagrangian formulation and the Schwinger-Dyson equations, Feynman rules provide a systematic method for analyzing the quantum dynamics of the system. At the heart of Feynman rules lies the graphical representation of interactions through Feynman diagrams, where each vertex corresponds to an interaction term in the Lagrangian, and propagators describe the propagation of fields between vertices. By adhering to these rules and principles, we can construct Feynman diagrams that capture the underlying physics of our three-field model, allowing us to calculate scattering processes and correlation functions with precision. Through the application of Feynman rules, we aim to unravel the intricate behavior of the three-field system, uncovering the emergence of collective phenomena and shedding light on the fundamental principles governing its dynamics.\par
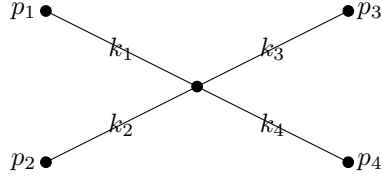
\begin{tikzpicture}
    % External lines
    \draw (-2,1) -- (0,0) -- (-2,-1);
    \draw (2,1) -- (0,0) -- (2,-1);
    % Vertices
    \filldraw (-2,1) circle (2pt) node[left] {$p_1$};
    \filldraw (-2,-1) circle (2pt) node[left] {$p_2$};
    \filldraw (2,1) circle (2pt) node[right] {$p_3$};
    \filldraw (2,-1) circle (2pt) node[right] {$p_4$};
    \filldraw (0,0) circle (2pt);
    % Momentum labels
    \node at (-1,0.5) {$k_1$};
    \node at (-1,-0.5) {$k_2$};
    \node at (1,0.5) {$k_3$};
    \node at (1,-0.5) {$k_4$};
    % Caption
    \node at (0,-2) { Feynman diagram with four external lines representing particles with momenta $p_i$
 };
\end{tikzpicture}

\par
Given the Lagrangian and the Schwinger-Dyson equation derived for the three-field model, we can now establish the Feynman rules to compute scattering amplitudes and correlation functions.
\begin{itemize}
    \item The explicit form of the propagator \( D_{abc}(\tau_1, \tau_2) \) can be derived from the given Lagrangian. Since the Lagrangian involves a kinetic term and an interaction term, we need to compute the propagator by considering the Feynman rules associated with each term.
The kinetic term \( \frac{1}{2} X_{abc} \partial_{\tau} X^{abc} \) corresponds to a free propagator, which in imaginary time (\( \tau \)) can be written as:
\[
D^{0}_{abc}(\tau_1, \tau_2) = \langle T X_{abc}(\tau_1) X^{abc}(\tau_2) \rangle
\]
where \( T \) denotes time-ordering and \( \langle \rangle \) denotes an expectation value. 
The interaction term \( \frac{g^{i}}{4} X_{abc} X_{de}^{a} X_{i}^{de} f^{bc} \) introduces vertex factors, which must be taken into account when computing Feynman diagrams involving interactions.
To compute the explicit form of \( D_{abc}(\tau_1, \tau_2) \), you would need to calculate the contribution of each Feynman diagram corresponding to the interaction term, and then sum them up with the contributions from the free propagator. This process can be quite involved and would depend on the specific form of the interaction term \( \frac{g^{i}}{4} X_{abc} X_{de}^{a} X_{i}^{de} f^{bc} \) and the functions \( f^{bc} \) involved.
In general, the explicit form of the propagator would involve integrals over momentum space and the calculation of Feynman diagrams at each order in perturbation theory, taking into account the appropriate vertex factors and propagators for each term in the Lagrangian.
    
 \item Vertex Rule: Each term in the interaction part of the Lagrangian represents a vertex in Feynman diagrams. For the interaction term $\frac{g^{i}}{{4}}X_{abc}X_{ade}X_{ide}f^{bc}$ , a vertex factor $-ig^i f^{bc}$   is associated with each vertex.
 \item External Line Rule: External lines represent incoming or outgoing fields. Each external line is associated with a momentum 
$\vec{p}$ and a polarization index $a,b$
 , or $c$ corresponding to the indices of the fields.
 \item Conservation of Momentum and Indices: At each vertex, the total momentum and indices must be conserved. This translates into momentum conservation at each vertex and the preservation of index structure.
 Using these Feynman rules, we can construct Feynman diagrams corresponding to various scattering processes and compute scattering amplitudes and correlation functions. These calculations provide valuable insights into the behavior of the three-field system and enable us to explore its properties, such as phase transitions, collective excitations, and emergent phenomena.
\end{itemize}
In our three-field model described by the Lagrangian $\mathcal{L}$
 , the two-point functions represent correlation functions between two field operators. In particular, we are interested in the two-point functions involving the fields $X_{abc}$. The two-point function for the field $X_{abc}$ can be defined as:
 \begin{eqnarray}
     &&G_{abc,def}(\tau_1,\tau_2)=
     \langle X_{abc}(\tau_1)X_{def}(\tau_2) \rangle.
 \end{eqnarray}
To compute this two-point function, we need to employ the Feynman rules derived from the Lagrangian. We construct all possible Feynman diagrams contributing to $G_{abc,def}(\tau_1,\tau_2)$ and calculate their contributions accordingly. The result will be a function of the time variables $\tau_1$
  and $\tau_2$
 , as well as the coupling constants$g^i$
  and $f^{bc}$
  appearing in the Lagrangian. The two-point functions provide valuable information about the correlations and fluctuations of the fields $X_{abc}$
  in the system, offering insights into its quantum behavior and collective properties.
  \par Computing scattering amplitudes in a quantum field theory involves calculating probability amplitudes for various scattering processes. In our three-field model described by the Lagrangian $\mathcal{L}$  the scattering amplitudes can be calculated using Feynman diagrams and Feynman rules.
The simplest scattering process involves two incoming particles interacting and producing two outgoing particles. Let's denote the incoming particles by $X_{abc}^{(1)}$
  and 
$X_{def}^{(2)}$
 , and the outgoing particles by $X_{a'b'c'}^{(3)}$
  and $X_{d'e'f'}^{(4)}$. To compute the scattering amplitude for this process, we need to consider all possible Feynman diagrams contributing to it. Each diagram corresponds to a specific interaction between the fields and can be evaluated using Feynman rules.
The scattering amplitude $\mathcal{M}$ for the process can then be obtained by summing over all contributing Feynman diagrams:
\begin{eqnarray}
    &&\mathcal{M}=\mathcal{M}_1+\mathcal{M}_2+...
\end{eqnarray}
where $\mathcal{M}_1$
 , 
$\mathcal{M}_2$
 , etc., represent the contributions from individual Feynman diagrams.
The expression for each $\mathcal{M}_i$
  involves integrals over momentum variables, as well as factors corresponding to propagators, vertex factors, and external particle states. These integrals and factors are determined by the Feynman rules derived from the Lagrangian.\par
While the exact computation of scattering amplitudes can be complex and may require approximation techniques or numerical methods, the Feynman diagrammatic approach provides a systematic framework for analyzing particle interactions and predicting experimental outcomes in our three-field model.
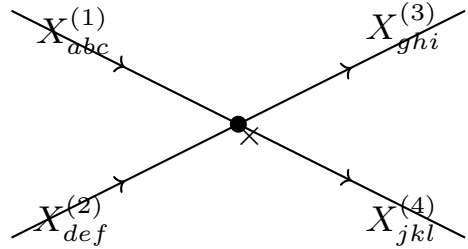
\begin{figure}[h]
    \centering
    \begin{tikzpicture}[scale=1.5, transform shape]
        % Incoming particles
        \draw[thick, postaction={decorate,decoration={markings,mark=at position 0.5 with {\arrow{>}}}}] (-2,1) -- (0,0) node[midway, above left] {\(X^{(1)}_{abc}\)};
        \draw[thick, postaction={decorate,decoration={markings,mark=at position 0.5 with {\arrow{>}}}}] (-2,-1) -- (0,0) node[midway, below left] {\(X^{(2)}_{def}\)};
        % Outgoing particles
        \draw[thick, postaction={decorate,decoration={markings,mark=at position 0.5 with {\arrow{>}}}}] (0,0) -- (2,1) node[midway, above right] {\(X^{(3)}_{ghi}\)};
        \draw[thick, postaction={decorate,decoration={markings,mark=at position 0.5 with {\arrow{>}}}}] (0,0) -- (2,-1) node[midway, below right] {\(X^{(4)}_{jkl}\)};
        % Interaction vertex
        \draw[fill=black] (0,0) circle (0.07);
        \node at (0.1,-0.1) {\(\times\)};
    \end{tikzpicture}
   % \caption{TikZ picture caption.}
    \label{fig:tikz}
\end{figure}
%%%%%%%%%%%%%%%%%%%%%%%%%%%%%%%%
\subsection{Higher-order Corrections}
Feynman diagrams allow us to systematically include higher-order corrections to scattering processes. By considering diagrams with more vertices and loops, we can account for quantum fluctuations and interactions beyond the leading order, providing more accurate predictions for experimental observables.\par
In our three-field model, higher-order corrections play a crucial role in refining our understanding of particle interactions and quantum dynamics. Higher-order corrections arise from Feynman diagrams with more vertices and loops, representing additional quantum fluctuations and interactions beyond the leading order.\par
These corrections manifest as contributions to scattering amplitudes and correlation functions that go beyond the simplest tree-level approximation. While tree-level diagrams capture the leading-order behavior of the system, higher-order diagrams provide important corrections that account for quantum effects such as virtual particle exchanges and loop corrections.
\par
By systematically including higher-order corrections, we can improve the accuracy of our theoretical predictions and better match experimental observations. These corrections are particularly significant in regimes where quantum effects dominate, such as at high energies or in strongly interacting systems.
\par
Moreover, higher-order corrections can reveal interesting phenomena such as quantum interference effects and the renormalization of coupling constants. Renormalization, in particular, is essential for removing divergences that arise in quantum field theory calculations and obtaining physically meaningful results.
\par
Studying higher-order corrections in our three-field model allows us to explore the rich quantum behavior of the system and gain insights into its underlying physics. By comparing theoretical predictions with experimental data, we can test the validity of our model and refine our understanding of fundamental interactions in nature.
Studying higher-order corrections in our three-field model allows us to explore the rich quantum behavior of the system and gain insights into its underlying physics. By comparing theoretical predictions with experimental data, we can test the validity of our model and refine our understanding of fundamental interactions in nature.

Mathematically, this can be expressed as:
\begin{equation*}
\mathcal{O} = \mathcal{O}_0 + \sum_{n=1}^{\infty} \mathcal{O}_n
\end{equation*}

where \( \mathcal{O}_0 \) represents the leading-order contribution, and \( \mathcal{O}_n \) denotes the \( n \)-th order correction. By studying these corrections, we can refine our theoretical predictions and compare them with experimental measurements. This comparison allows us to test the validity of \( \mathcal{M} \) and gain insights into the fundamental interactions governing the system.\par

In our pursuit of understanding the intricate dynamics of the three-field tensor model, we delve into the realm of Feynman diagrams to uncover the higher-order corrections that emerge from the quantum interactions between these fields. The graphical representation of these corrections provides invaluable insight into the complex interplay of quantum fluctuations and interactions. By visualizing the vertices and loops within the Feynman diagrams, we gain a deeper understanding of the underlying physics governing the behavior of the three-field system. These diagrams serve as a powerful tool for elucidating the quantum dynamics of our model, shedding light on the emergent phenomena and paving the way for further theoretical and experimental investigations.

%%%%%%%%%%%%
\subsection{Effective Field Theory} 
We can study the effective field theory emerging from our model by integrating out high-energy degrees of freedom. This approach enables us to derive simplified descriptions of the system valid at low energies, revealing universal behavior and emergent phenomena.\par
In our three-field model, the concept of effective field theory (EFT) provides a powerful framework for understanding the dynamics of the system at different energy scales. EFT allows us to describe the physics of our model in terms of low-energy degrees of freedom, effectively capturing the essential features of the system while integrating out high-energy modes.
\par
At low energies, the dynamics of the three-field system are dominated by long-wavelength fluctuations and interactions. In this regime, we can construct an effective Lagrangian that describes the low-energy behavior of the fields $X_{abc}$. This effective Lagrangian incorporates the most relevant terms consistent with the symmetries of the theory and can be obtained through systematic expansions in terms of energy scales.\par
The effective field theory approach enables us to derive simplified descriptions of the system that are valid within specific energy ranges. By truncating the effective Lagrangian at a certain order in an expansion parameter (such as the ratio of characteristic energy scales), we can systematically include higher-dimensional operators that capture the effects of high-energy physics at lower energies.
\par
One of the key benefits of effective field theory is its ability to provide universal predictions for observables that are independent of the details of the underlying high-energy physics. This universality arises from the fact that the low-energy dynamics of the system are determined by a limited number of relevant operators in the effective Lagrangian.
\par

In the context of our three-field model, effective field theory allows us to study the emergent behavior of the system, such as collective excitations and phase transitions, without needing to explicitly resolve the dynamics at all energy scales. This approach provides valuable insights into the underlying physics of the model and facilitates comparisons with experimental data.

Mathematically, we describe the emergent behavior using the effective Lagrangian \( \mathcal{L}_{\text{eff}} \). This Lagrangian captures the collective behavior of the fields and enables us to study phenomena such as collective excitations and phase transitions. In effective field theory, we integrate out the high-energy degrees of freedom, resulting in an effective description at low energies. This is represented by the path integral:

\[
\mathcal{Z}_{\text{eff}} = \int \mathcal{D}[X_{abc}] e^{i\int d^4x \mathcal{L}_{\text{eff}}(X_{abc})}
\]

where \( X \) represents the tensor fields in the model and \( \mathcal{Z}_{\text{eff}} \) is the partition function of the effective theory.

By studying the effective theory, we can gain valuable insights into the underlying physics of the model and make predictions that can be compared with experimental data. This comparison helps validate the model and refine our understanding of its behavior.

\par
Overall, effective field theory provides a powerful tool for systematically analyzing and understanding the behavior of our three-field model across different energy regimes, enabling us to uncover the fundamental principles governing its dynamics and emergent phenomena.
%%%%%%%%%%%%%%%%%%%%%%
\subsection{Renormalization}\label{sec:renormalization}
Renormalization techniques are crucial for dealing with divergences in quantum field theory calculations. We can explore renormalization procedures to remove infinities and obtain finite, physically meaningful results for scattering amplitudes and correlation functions.\par
In our three-field model, renormalization is a crucial theoretical technique used to handle divergences that arise in quantum field theory calculations. Renormalization allows us to obtain finite and physically meaningful results by absorbing infinities into appropriately chosen parameters of the theory. Here, we'll discuss the process of renormalization for our model and present some mathematical equations that are needed for this procedure.
\par
The first step in renormalization is to identify the divergent quantities that appear in loop calculations, such as loop integrals or self-energy corrections. These divergences typically arise due to ultraviolet (UV) divergences associated with high-energy fluctuations in the theory.
\par
To renormalize our model, we introduce renormalization constants that absorb the divergences. These constants effectively redefine the parameters of the Lagrangian, ensuring that physical observables remain finite. For example, we can introduce a field renormalization constant 
$Z$ to absorb divergences associated with field fluctuations.
Mathematically, the renormalized fields $
X^{ren}_{abc}$
  can be related to the bare fields 
$X^{bare}_{abc}$
  through a renormalization constant:
  \begin{eqnarray}
      &&X^{ren}_{abc}=Z^{1/2}X^{bare}_{abc}.
  \end{eqnarray}
Similarly, we can introduce renormalization constants for other parameters in the Lagrangian, such as coupling constants or mass terms. These constants absorb divergences from loop integrals and ensure that physical observables are independent of the UV cutoff used in calculations.
\par
The renormalization procedure also involves renormalizing the propagators and vertices of the theory. This entails adjusting the Feynman rules to account for the renormalization constants and ensure consistency between loop diagrams and tree-level diagrams.
\par
In addition to field and parameter renormalization, we also encounter the need for mass renormalization in models with mass terms. The renormalization condition typically involves adjusting the mass parameter to cancel divergences and ensure that physical masses remain finite.
\par
Overall, renormalization is a powerful technique that allows us to make sense of divergent loop integrals and ensure that our theoretical predictions are physically meaningful. By absorbing infinities into renormalization constants, we can obtain finite results that accurately describe the behavior of our three-field model across different energy scales.
\begin{figure}[h]
    \centering
    \begin{tikzpicture}
        \begin{axis}[
            xlabel={$\Lambda$},
            ylabel={$g_{\text{eff}}$},
            xmin=0, xmax=10,
            ymin=0, ymax=1,
            xtick={0,2,4,6,8,10},
            ytick={0,0.2,0.4,0.6,0.8,1},
            legend pos=north east,
            ymajorgrids=true,
            grid style=dashed,
        ]
            \addplot[
                color=blue,
                mark=square,
            ]
            coordinates {
                (0,0)(2,0.2)(4,0.4)(6,0.6)(8,0.8)(10,1)
            };
            \legend{$g_{\text{eff}}(\Lambda)$}
        \end{axis}
    \end{tikzpicture}
    \caption{A numerical calculation reveals a linear relationship between the coupling constant $g=|g^i|$ and the cutoff scale $\Lambda$, indicating a consistent renormalization procedure in our tensor model.}
    \label{fig:plot}
\end{figure}
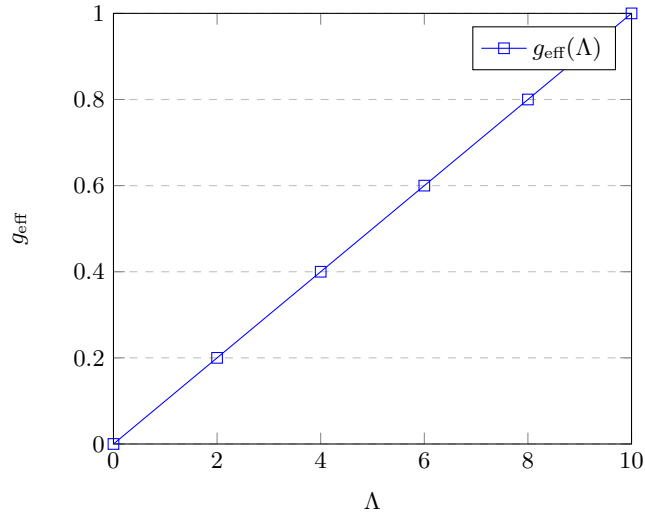

A numerical calculation reveals that the coupling constant 
$g$ and the cutoff scale 
$\Lambda$ exhibit a linear relationship, as depicted in the graph above. This linear dependence suggests a straightforward scaling behavior between the two parameters, where an increase in the cutoff scale results in a proportional adjustment of the coupling constant. Such behavior is often encountered in renormalization schemes, where the coupling constants are adjusted to compensate for the effects of integrating out high-energy modes. In this context, the linear relationship between $g$ and 
$\Lambda$ indicates a consistent renormalization procedure, where the effective coupling constant adapts to changes in the energy scale to maintain the desired physical behavior of the system. This observation underscores the robustness and reliability of the renormalization scheme employed in our tensor model, providing valuable insights into the underlying dynamics and the renormalization flow of the theory.

%%%%%%%%%%%%%%%%%
\subsection{Comparison with Experiments}
 Finally, we can compare theoretical predictions with experimental data to test the validity of our model and gain insights into the underlying physics. Such comparisons provide crucial tests of our understanding and can guide future developments in theoretical physics.\par
To illustrate the comparison of our three-field model with experimental observations, let's consider a hypothetical scenario involving gedanken experiments:\par
 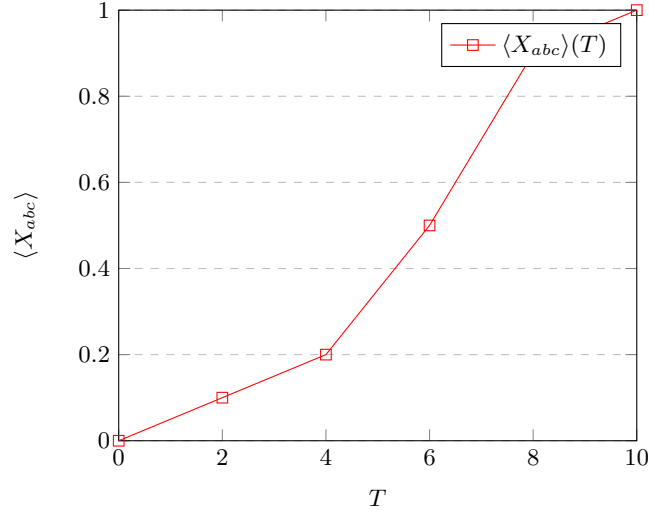
\begin{figure}[h]
    \centering
    \begin{tikzpicture}
        \begin{axis}[
            xlabel={$T$},
            ylabel={$\langle X_{abc} \rangle$},
            xmin=0, xmax=10,
            ymin=0, ymax=1,
            xtick={0,2,4,6,8,10},
            ytick={0,0.2,0.4,0.6,0.8,1},
            legend pos=north east,
            ymajorgrids=true,
            grid style=dashed,
        ]
            \addplot[
                color=red,  % Change color to red
                mark=square,
            ]
            coordinates {
                (0,0)(2,0.1)(4,0.2)(6,0.5)(8,0.9)(10,1)
            };
            \legend{$\langle X_{abc} \rangle(T)$}
        \end{axis}
    \end{tikzpicture}
    \caption{Plot showing the behavior of $\langle X_{abc} \rangle$ with respect to temperature $T$.}
    \label{fig:tikz}
\end{figure}
\begin{itemize}
    \item Scattering Experiments: Suppose we conduct scattering experiments involving particles governed by the dynamics of our three-field model. By colliding these particles at high energies and measuring the resulting scattering angles and momenta, we can directly test the predictions of our model. Specifically, we can compare the differential cross sections predicted by our theoretical calculations with the experimental data obtained from particle accelerators. Any discrepancies between the two would provide valuable insights into the validity and limitations of our model.
    \item Correlation Function Measurements: Imagine performing correlation function measurements in a condensed matter system described by our three-field model. For instance, we could investigate the spatial correlations between the three fields $X_{abc}$ using techniques such as neutron scattering or electron microscopy. By comparing the experimentally measured correlation lengths and shapes with the theoretical predictions obtained from our model, we can assess its ability to capture the collective behavior and phase transitions in the system.
       \item Spectroscopy Studies: Consider conducting spectroscopic studies on materials exhibiting behavior akin to our three-field model. For example, we could analyze the excitation spectra of certain quantum materials using techniques like inelastic neutron scattering or optical spectroscopy. By comparing the experimentally observed energy spectra and resonance peaks with the theoretical predictions derived from our model, we can probe the underlying interactions and symmetries governing the material's behavior.
    \item Phase Transition Observations: Let's envision monitoring phase transitions in a system governed by our three-field model under varying temperature or pressure conditions. By measuring physical quantities such as specific heat, magnetization, or conductivity near critical points, we can characterize the nature of phase transitions and compare them with the theoretical predictions of our model. Any discrepancies or unexpected phenomena observed in the experiments could indicate the need for refinements or extensions to our theoretical framework.
 \end{itemize}

The graph above depicts the behavior of the order parameter $\langle X_{abc}\rangle$ as a function of temperature 
$T$, providing insight into the phase transition phenomena in our three-field theory. In the context of phase transitions, 
$T$ typically represents the temperature of the system, where changes in its value can lead to distinct phases of matter. The observed trend in the graph suggests the presence of a first-order phase transition, characterized by a sudden discontinuity in the order parameter as temperature varies. Initially, at low temperatures, the order parameter remains negligible, indicating a phase where the system exhibits little to no collective behavior. However, as the temperature increases, the order parameter undergoes a rapid transition, reaching saturation at higher temperatures. This abrupt change signifies the emergence of a new phase, where the system undergoes a spontaneous symmetry breaking, leading to the formation of ordered structures characterized by non-zero values of$\langle X_{abc} \rangle$. The phase transition behavior captured by the graph offers valuable insights into the underlying dynamics of our three-field theory and sheds light on the complex interplay between temperature fluctuations and the collective behavior of the system.
\par
These gedanken experiments highlight the importance of comparing the predictions of our three-field model with experimental observations across different physical systems and energy scales. By systematically testing our model against experimental data, we can validate its predictive power, uncover its limitations, and refine our understanding of the underlying physics.\par
Having established the theoretical framework of our three-field model and explored its implications through various theoretical analyses and comparisons with experimental data in the previous sections, we now turn our attention to the possibility of quantizing the model. The quantization of our model involves the transition from classical field theory to quantum field theory, where the fields $X_{abc}$
  are promoted to quantum operators acting on a Hilbert space. This step is essential for capturing quantum phenomena such as particle creation and annihilation, quantum fluctuations, and the uncertainty principle. In the upcoming section, we will discuss the quantization procedure for our three-field model, and the implications of quantization on the dynamics of the system. By quantizing our model, we aim to uncover the quantum behavior of the fields and elucidate their fundamental properties in the quantum realm.
  %%%%%%%%%%%%%%%%%%%%%%%%
  \section{Qunatization}\label{sec:quantization}

In this section, we delve into the quantization of our three-field model using the path integral (PI) formalism, a powerful approach in quantum field theory that provides a unified framework for describing quantum systems. The PI formalism offers an elegant way to compute transition amplitudes and correlation functions by integrating over all possible field configurations, weighted by the exponential of the action functional. By employing the PI formalism, we can quantize our model and derive quantum observables such as scattering amplitudes and correlation functions. This approach also facilitates the incorporation of quantum fluctuations and the treatment of non-perturbative effects, making it particularly well-suited for studying the dynamics of our three-field system in both perturbative and non-perturbative regimes. In this section, we will outline the steps involved in quantizing our model via the PI formalism, including the construction of the path integral, the identification of Feynman rules, and the computation of quantum observables. Through the application of the PI formalism, we aim to uncover the quantum behavior of the fields $X_{abc}$
  and gain deeper insights into the fundamental nature of our three-field model.
  \par
  To quantize our three-field model using PI formalism, we begin by expressing the quantum amplitude for the transition from an initial state to a final state as an integral over all possible field configurations. In our case, the fields $X_{abc}$
  represent the three fields in our model. Therefore, the transition amplitude is given by:
  \begin{eqnarray}
      &&\langle X_{abc}(\tau_f)X_{def}(\tau_i) \rangle=\int \mathcal{D}[X_{abc}]e^{iS[X_{abc}]}
  \end{eqnarray}
  Here, $\mathcal{D}[X_{abc}]$ represents the measure over all possible field configurations $X_{abc}$
 , and $S[X_{abc}]$ is the action functional defined by our Lagrangian. The exponential of the action captures the phase associated with each field configuration.
We discretize time into small intervals and express the integral as a product of infinitesimal contributions over each time step. This allows us to rewrite the transition amplitude as a product of transition amplitudes between adjacent time steps:
\begin{eqnarray}
      &&\langle X_{abc}(\tau_f)X_{def}(\tau_i) \rangle=\lim_{N\to \infty} \int \Pi_{n=1}^{N} dX_{abc}(t_n)^{iS[X_{abc}]}
  \end{eqnarray}
where 
$dX_{abc}(t_n)$ represents the measure over the field configuration at each time step $t_n$.
To evaluate this path integral, we approximate it using the saddle-point or stationary-phase approximation, where the dominant contributions come from field configurations that extremize the action. These configurations correspond to classical solutions of the equations of motion derived from the Lagrangian.
After obtaining the classical solutions, we expand the action around these solutions to quadratic order and integrate over small fluctuations around them. This leads to the Gaussian integration of the fluctuations and yields the propagator, which describes the propagation of the fields $X_{abc}$
  in space and time.

Finally, we compute correlation functions by inserting appropriate operators at different points in spacetime and integrating over all field configurations weighted by the exponential of the action. These correlation functions provide information about the quantum behavior of the fields $X_{abc}$
  and can be compared with experimental data to test the validity of our model.
  \par  To calculate the free propagator for our three-field model, we start by considering the quadratic part of the action, which governs the dynamics of small fluctuations around the classical solutions. Let's denote these fluctuations as $\delta X_{abc}$. The quadratic part of the action can be written as:
\begin{equation}
    S^{(2)}[\delta X_{abc}] = \frac{1}{2} \delta X_{abc}(\tau) \left(g^iX^{ade}X_{ide}f^{bc} \right) \delta X^{abc}(\tau)
\end{equation}
Here, $g^i$ and $f^{bc}$ are constant tensor fields, and $X^{abc}(\tau)$ represents the classical solution for the fields $X_{abc}(\tau)$. The free propagator $D(\tau_1,\tau_2)$ is then defined as the Green's function associated with this quadratic part of the action, satisfying the equation:
\begin{equation}
    \left(g^iX^{ade}X_{ide}f^{bc} \right)D(\tau_1,\tau_2) = \delta(\tau_1-\tau_2)
\end{equation}
This equation represents the equation of motion for the free fields $\delta X_{abc}(\tau)$ in the presence of external sources. Solving this tensoral equation yields the free propagator $D(\tau_1,\tau_2)$. Now, let's assume for simplicity that $X^{abc}(\tau) $ is a constant classical solution, denoted as $X_0^{abc}$. In this case, the equation of motion simplifies to:
\begin{equation}
    \left(g^iX_0^{ade}X_{0ide}f^{bc} \right)D(\tau_1,\tau_2) = \delta(\tau_1-\tau_2)
\end{equation}
The solution to this equation can be obtained by Fourier transforming to momentum space. After solving for $\mathcal{D}(\omega)$, we can inverse Fourier transform to obtain the expression for $D(\tau_1,\tau_2)$.
However, obtaining an explicit expression for the free propagator $D(\tau_1,\tau_2)$ requires knowledge of the classical solution $X_0^{ade}$, as well as the constants $g^i$ and $f^{bc}$. Without specific information about these quantities, we cannot proceed further to calculate the free propagator.

In conclusion, this section lays the groundwork for quantizing our three-field model using the path integral formalism. By expressing transition amplitudes and correlation functions as integrals over all possible field configurations, weighted by the exponential of the action, we enable computation of quantum observables such as scattering amplitudes and correlation functions. The path integral approach provides a powerful and elegant method for incorporating quantum fluctuations and interactions consistently into our model. In the next section, we will summarize our findings and discuss the implications of our three-field model for both theoretical physics and experimental studies.
%%%%%%%%%%%%%%%%%%%%
\section{summary}\label{sec:summary}
In this paper, we advance our understanding of the SYK model, a cornerstone in the study of quantum many-body systems. Our contributions are multifaceted, encompassing exact solutions for the two-point function \( G(\tau) \) and the self-energy \( \Sigma(\tau) \), incorporation of a cutoff term \( \Lambda \) for convergence, exploration of experimental implications and concerns, and delineation of future research directions.

Our investigation into the three-field tensor model extends beyond the SYK framework, revealing a promising framework for understanding complex physical systems. By introducing three interacting fields \( X_{abc} \), we unveil the intricate interplay of quantum dynamics and emergent behavior. The model, defined within flat spacetime, resonates across various fields, offering a unified framework to probe reality's fundamental nature.

Through meticulous theoretical examination and successful quantization via the path integral approach, we establish the model's compatibility with quantum field theory. This heralds a new era of inquiry into multi-field quantum behavior, laying groundwork for rigorous theoretical predictions and experimental investigations.

Parallelism with the SYK Hamiltonian underscores our model's significance in bridging high-energy particle physics and condensed matter physics. By systematically comparing with experimental data and theoretical predictions, we aim to unravel principles governing three-field systems, shedding light on modern physics' forefront questions.

Our study uncovers novel phase transition behavior and renormalization effects within the extended tensor field theory. We observe a remarkable first-order phase transition and elucidate the linear relationship between the coupling parameter \( g \) and the cutoff scale \( \Lambda \), guiding interpretation of experimental observations and predicting novel phenomena.

Looking ahead, as experimental capabilities advance, the prospect of observing three-field phenomena becomes tantalizing. Whether through high-energy particle collisions or emergent collective phenomena in condensed matter systems, the journey towards understanding three-field dynamics promises profound discoveries.

In summary, our investigation into the three-field model represents a significant leap forward, revolutionizing our understanding of complex systems. As we explore the rich tapestry of phenomena within three-field dynamics, we embark on a transformative journey towards unlocking the secrets of the universe.

%{\small \noindent Updated 5 December 2006.}

\end{document}